\documentclass[10pt,twoside]{article}

\usepackage{asp2006}
\usepackage{epsf}
\usepackage{psfig}
\usepackage{lscape}
\usepackage{graphicx}

\begin{document}
\title{Constraining Galaxy Evolution With Bulge-Disk-Bar Decomposition}   
\author{Tim Weinzirl\altaffilmark{1}, Shardha Jogee\altaffilmark{1}, Fabio D. Barazza\altaffilmark{2}}   

\altaffiltext{1}{The University of Texas, Department of Astronomy, 1 University Station, C1400, Austin, Texas 78712-0259}
\altaffiltext{2}{Laboratoire d'Astrophysique, \'Ecole Polytechnique F\'ed\'erale de Lausanne (EPFL), Observatoire, 1290 Sauverny, Switzerland}

\begin{abstract} 

Structural decomposition of galaxies into bulge, disk, and bar components is
important to address a number of scientific problems.  Measuring bulge, disk,
and bar structural parameters will set constraints on the violent and secular
processes of galaxy assembly and recurrent bar formation and dissolution
models.  It can also help to quantify the fraction and properties of bulgeless
galaxies (those systems having no bulge or only a relatively insignificant
disky-pseudobulges), which defy galaxy formation paradigms requiring almost
every disk galaxy to have a classical bulge at its core.

We demonstrate a proof of concept and show early results of our ongoing
three-component bulge-disk-bar decomposition of NIR images for a sample of
three complementary samples spanning different epochs and different
environments (field and cluster).
In contrast to most early studies,
which only attempt two-component bulge-disk decomposition, we fit three
components using GALFIT: a bulge, a disk, and a bar. We show that it is
important to include the bar component, as this can significantly lower the
bulge-to-total luminosity ratio (\emph{B/T}), in many cases by a factor of two
or more, thus effectively changing the Hubble type of a galaxy from early to
late.
\end{abstract}

\section{Introduction}
The formation of galaxies is a classic problem in astrophysics.
Contemporary galaxy formation models combine the
well-established Lambda-Cold Dark Matter (LCDM) cosmology, which describes
behavior of dark matter on very large scales, with baryonic physics to model
galaxy formation. In the early Universe, pockets of dark matter decoupled from
the Hubble flow, collapsed into virialized halos, and then clustered
hierarchically into larger structures. Meanwhile, gas aggregated in the
interiors of the halos to form rotating disks, which are the building blocks
of galaxies (Navarro \& Steinmetz, 2002; Cole et al. 2000). 
Such disks were destroyed during mergers of their parent halos,
leaving behind classical de Vaucouleurs bulges. Spiral disk galaxies formed
subsequently as gaseous disks accreted around spheroids (Burkert \& Naab, 2004).

Troubling inconsistencies exist between real galaxies and LCDM models of
galaxy formation. One issue is the angular momentum problem; simulated galaxy
disks have smaller scalelengths and, therefore, less specific angular momentum
than their counterparts in nature (D'Onghia \& Burkert, 2006). A second problem
is the severe under prediction in the fraction of galaxies with 
low bulge-to-total mass ratio ($B/T<$0.2) and of so-called bulgeless 
galaxies, which lack a classical bulge. Simulated spiral galaxies feature 
prominent classical bulges in their cores. Such predictions are in glaring
contradiction with emerging observations that suggest 15-20\% of disk
galaxies out to z$\sim$0.3 are bulgeless (Kautsch et al. 2006; Barazza et al. 2007)

There are many unanswered questions about the assembly of bulges, the
distribution of $B/T$, and the properties of so-called bulgeless galaxies
with low $B/T$. How do  properties, such as disk scalelengths, mass,
kinematics, colors, and star formation histories vary across galaxies
of different $B/T$, ranging from bulge-dominated systems to quasi-bulgeless
systems?  Are quasi-bulgeless
systems confined to low mass systems with high specific star formation rates,
while classical bulges populate high mass systems?
How do the fraction,  mass function, and structural properties of
galaxies with different $B/T$ vary across environments with different
large-scale cosmological overdensities?  If environment plays a central part
in suppressing bulge formation, then differences would be expected
in the properties of bulgeless galaxies in different environments,
such as field versus dense galaxy clusters.
How does the frequency and properties of galaxies with low $B/T$ as a
function of redshift over $z=0.2-0.8$ compare to the recently reported
merger history of galaxies over this epoch (Jogee et al 2007)?
Answering these questions will help us to understand the reasons behind
the apparent failure of LCDM galaxy formation models, and shed light on
how galaxies assemble.

Progress is possible by observationally constraining properties of enigmatic
bulgeless galaxies.  A powerful technique for measuring the structural
properties (e.g. scalelengths, S\'ersic indexes, \emph{B/T}) of galaxies is the
decomposition of the 2D light distribution into separate structural components
with GALFIT (Peng et al. 2002). Most earlier work has only performed 2D
bulge-disk decomposition, but because late-type spirals have been shown to
have higher optical bar fractions than early-type galaxies
(Barazza et al. 2007), it is important to include the bar when analyzing
disk-dominated systems. Bars can contain a significant fraction of light, so
failure to account for bars could lead to inflated \emph{B/T}
(Laurikainen et al. 2006).

\section{Methodology and Samples}
We perform three-component decomposition of the 2D galaxy light distribution,
while taking the PSF into account, with GALFIT.  
Since GALFIT utilizes a non-linear least squares algorithm, initial guess 
parameters are required for each component GALFIT attempts to fit.  
While reasonable initial guesses can be generated by inspection in 
many cases with common tools (e.g. IRAF), this is time-consuming and 
inefficient for large samples.
In practice, we break three-component decomposition into three separate 
invocations of GALFIT.

We first perform one and two-component fits to constrain the bulge and disk
parameters.  The single-component fit models the entire galaxy with only a
S\'ersic bulge component.  In addition to constraining the bulge structural
parameters, the total
luminosity of the object is also determined.

A two-component fit, consisting of a S\'ersic bulge and exponential disk,
is then made based on the output of the previous fit.
If GALFIT is allowed to do an unconstrained two-component bulge-disk fit
in a strongly barred galaxy,  it will often try to fit the bar by
artificially stretching the disk along the bar PA. In order to get
physically meaningful two-component fits, we therefore constrain the fit by
fixing the position  angle and axis ratio (b/a) of the outer disk
to values pre-determined by fitting an ellipse to the outermost disk
isophote.

Finally, a  three-component bulge-bar-disk fit is performed, using
the two-component fits as initial guesses for the bulge parameters,
and fixing the disk b/a and PA as before.
Bars are modeled with elongated, low-index S\'ersic components
using  initial guesses for  the size and  position angle estimated
from the images.  All objects are subjected to the 
three-component fits, regardless of whether they appear by eye 
to possess a bar.  If there is independent evidence for an AGN or nuclear
cluster, a point source is fitted as a fourth component.

In order to decide which of the two or three-component fit is better, a 
number of criteria are used.  1)~If the one or two-component residuals 
show a bar 
signature that is removed in the three-component residual, then the three
component fit is favored; 2)~Structural parameters (scalelength, S\'ersic 
index, b/a) of the bar fit must well behaved;  3)~Visual evidence of a 
strong bar in the input images favors the three-component fit. Weak 
bars are often not visually prominent, but for such bars, the changes
in the disk or bulge parameters, between the two and three component 
fits, are small; (4)  In addition, we test the robustness of the 
three-component solution by varying the initial guesses to check 
that the same solution is converged upon.

In order to address the questions outlined in \S1, we are applying
the three-component decomposition to three complementary samples, which span
different epochs and different environments (field and cluster):
(1)  a  $z\sim$~0 sample of $\sim$~200 galaxies with Hubble types S0 to Sm
drawn from the OSU Bright Spiral Galaxy Survey (OSUBSG) (Eskridge et
al. 2002) and UKIDSS (McLure et al. 2006); (2)  a sample of galaxies
in the  dense environment of the Coma cluster from our ACS
Treasury survey (Carter et al 2007); and a sample of early disk 
galaxies out $z\sim$~2  with deep NICMOS imaging (180 orbits).

\section{Preliminary Findings}
For the two-component fits, we have  performed consistency checks
by testing our decomposition on samples of galaxies with
published results from the Millennium Galaxy Catalog
(Driver et al. 2007) and 
the New York University Value-Added Catalog
(Blanton et al. 2005)
for the Sloan Digital Sky Survey.

For three-component fits, we have performed similar tests on a few
galaxies drawn from small samples  with published three-component
bulge-bar-disk fits, drawn from Laurikainen et al. (2006) and
Reese et al. (2007).

An example of our method is presented in Figure 1, which illustrates
the complete three-step decomposition for NGC 4643. 
We now summarize our preliminary findings:
\begin{enumerate}
 \item Luminosity is conserved between the two and three-component fits.
 \item Modeling the bar in the three-component fits forces a reshuffling of 
    luminosity.  Generally, the bulge declines in luminosity,
    whereas light can be taken from, or added back, to the disk.
    The reshuffling of light
    occurs because the two-component model adjusts the bulge and disk 
    accordingly to compensate for the bar, which can include artificially
    elongating and brightening the bulge.  Accounting for the bar returns the
    bulge and disk parameters closer to their true values.
 \item Inclusion of the bar can reduce bulge fractional luminosity \emph{B/T}
    by a factor of two or more.  Larger changes in bulge luminosities (a factor
    of 10 or more) occur
    in cases where a prominent bar influences the two-component fit to very much
    overstate the bulge luminosity.  The bulge-disk fits in such extreme
    cases underscore the importance of
    including the bar in 2D luminosity decomposition.
 \item The scalelength of the disk is generally unchanged by including 
    the bar.  However in a few cases, the two-component disk structure 
    can be erroneous, as in the case of NGC 4643, shown in Figure 1.
\end{enumerate}
We have provided a proof of concept of our ongoing three-component 
bulge-disk-bar decomposition with GALFIT.  We are optimistic about our
on-going work, which will be described in Weinzirl et al. 2008 (in prep).

\begin{figure}[!h]
\begin{center}
\includegraphics[angle=0,scale=0.7]{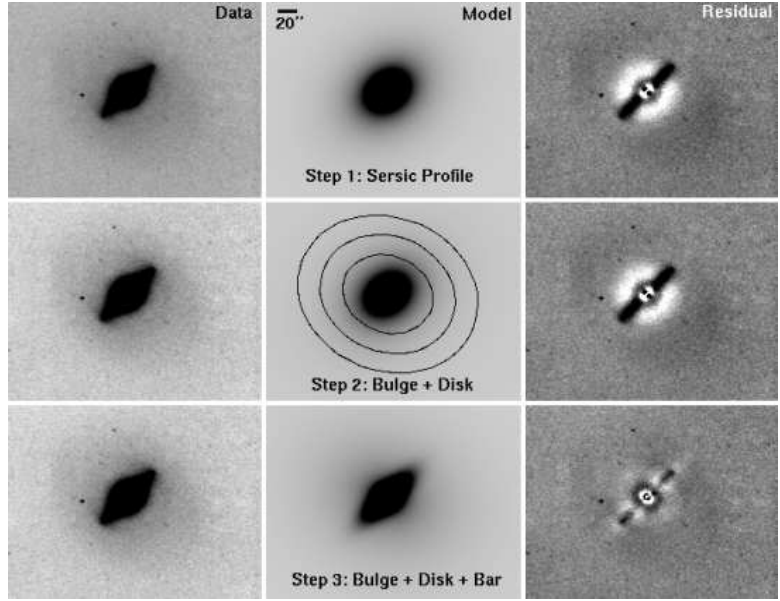}
\caption{Shown is the complete three-step decomposition for NGC 4643.  From top
to bottom, the rows show the fits from the one, two, and 
three-component decompositions. 
The residuals for the one and two-component fit show   a distinct bar 
signature. In Step 2, the fitted disk has an unphysically large 
scalelength ($335\arcsec$) that does not match the galaxy. Due to 
its resulting low surface brightness, the fittted disk is hard to 
see, and ellipses are drawn to show its PA and b/a. In Step 3, 
the addition of the bar component restores the disk scalength
to a reasonable value. The fit parameters are presented in Table 1.}
\end{center}
\end{figure}

\begin{table}[!h]
\caption{Fit parameters for NGC 4643}
\centering
\begin{tabular} {c c c c c c c}
\hline\hline
Fit   &      &$r_e$ or h ($\arcsec$) &  n    &  b/a  & Position Angle    &Fractional light \\
Step 1&Sersic &27.90               &  4.44 &  0.80 & -51.08  &100\%\\
\hline
Step 2&Bulge &23.86               &  4.16 &  0.80 & -51.08 &34.6\%\\
      &Disk  &335.88              &  1.0  &  0.84 & 66.94  &65.4\% \\
\hline
Step 3&Bulge &5.43                &  2.53 &  0.90 & 60.52  & 25.0 \% \\
      &Disk  &48.22               &  1.0  &  0.84 & 66.94  & 54.1 \% \\
      &Bar   &21.30               &  0.62 &  0.37 & -45.84 & 20.9 \% \\
\hline
\end{tabular}
\end{table}

\acknowledgements 

TW and SJ acknowledge support from NSF grad AST-0607748, LTSA grant NAG5-13063,
and HST-GO-10861 from STScI, which is operated by AURA, Inc., for NASA, under
NAS5-26555.

\end{document}